\documentclass[aps,twocolumn,pre,showpacs,floatfix,superscriptaddress]{revtex4}

\usepackage{amsmath,fixmath}
\usepackage{graphicx}
\usepackage{amssymb}
\usepackage{bm}
\begin{document}

\newcommand{\itsf}[1]{\textsf{\itshape #1}}

\title{Mean-value identities as an opportunity for Monte Carlo error reduction}

\author{L.A.~Fernandez} \affiliation{Departamento
  de F\'\i{}sica Te\'orica I, Universidad
  Complutense, 28040 Madrid, Spain.} 
  \affiliation{Instituto de Biocomputaci\'on y
  F\'{\i}sica de Sistemas Complejos (BIFI), Zaragoza, Spain.}

\author{V.~Martin-Mayor} \affiliation{Departamento de F\'\i{}sica
  Te\'orica I, Universidad Complutense, 28040 Madrid, Spain.} 
\affiliation{Instituto de Biocomputaci\'on y
  F\'{\i}sica de Sistemas Complejos (BIFI), Zaragoza, Spain.}

\date{\today}

\begin{abstract}
  In the Monte Carlo simulation of both Lattice field-theories and of
  models of Statistical Mechanics, identities verified by exact
  mean-values such as Schwinger-Dyson equations, Guerra relations,
  Callen identities, etc., provide well known and sensitive tests of
  thermalization bias as well as checks of pseudo random number
  generators. We point out that they can be further exploited as {\em
    control variates} to reduce statistical errors.  The strategy is
  general, very simple, and almost costless in CPU time.
  The method is demonstrated in the two dimensional Ising model at
  criticality, where the CPU gain factor lies between 2 and 4.
\end{abstract}
\maketitle 

\section{Introduction}
Monte Carlo simulation~\cite{MONTEBOOK,SOKALLECTURE} is one of the handful of
general methods in the theoretical physicist's toolbox that can be
applied to nonperturbative problems. In spite of this, it is a very
inefficient method: the computational effort needed to get yet another
decimal significant figure grows by a factor of 100.

Yet, there are alternatives to brute force when more accuracy is
needed.  A classical strategy consists in looking for statistical
estimators of the sought quantities, which have the same expectation
value as the commonly used naive estimators, but a reduced variance.
The {\em multihit} method~\cite{MULTIHIT} (and later
developments~\cite{LUSCHER}) for the Polyakov loop in Lattice QCD is a
conspicuous example of such an improvement.  Now, the numerical error
is proportional to the square root of variance for the considered
estimator. It follows that reducing the variance by a factor of two
reduces as well in the same factor the numerical effort needed to
achieve the desired statistical accuracy. Even a modest factor of
variance reduction can be a significant improvement: the CPU time
needed in application to Lattice Gauge Theory or to Condensed Matter
Physics (think for instance of Spin-Glass
simulations~\cite{JANUS-PRL}) oftenly lies in the range 10--$10^4$
processor years.

Here we propose a general road to variance-reduction based in known
identities between exact mean-values. In spite of its usefulness, this
strategy, known as {\em control variates} in the mathematical
literature~\cite{HAMMERSLEY,RUBINSTEIN}, is still not commonly used in
the framework of Monte Carlo simulations in Physics (at the practical level, it requires
only standard Monte Carlo data-analysis tools).  In fact, it is fairly
common to find in Field Theory or in Statistical Mechanics that a
particular linear combination of non-trivial expectations values
vanishes exactly (we provide below specific examples). There are
different ways of finding such identities: Schwinger-Dyson equations
exploit invariances of the integration measure~\cite{SCHWINGER-DYSON};
Callen identities are derived by integrating in the functional
integral some variable while holding fixed all others~\cite{CALLEN}
(multihit operators~\cite{MULTIHIT,LUSCHER} belong to this category);
Guerra relations are somehow specific to disordered
systems~\cite{GUERRA}; in models where a cluster method
works~\cite{SWENDSEN-WANG} cluster estimators with the same
expectation value than their spin counterparts can be found (see
e.g.~\cite{SALAS-SOKAL} and references therein). It is fair to say
that, for any problem amenable to a path-integral formulation, each of
the above strategies will provide at least one identity: the vanishing
of a precise linear combination of expectation values of non-trivial
observables.

Researchers performing Monte Carlo simulations are acutely aware of
the advantages provided by mean-value identities. If the numerically
obtained expectation values do not verify them within errors, this
will most probably be due to a thermalization bias~\cite{APE-SG,YOUNG}
or to a failure of the used Pseudo Random Number Generator~\cite{SD}
(or to a programming bug!). We remark here that mean-value identities provide
as well statistical estimators with reduced variance. The method is
exemplified in the standard benchmark of the two-dimensional Ising model
at its critical point.

We note finally that in previous work~\cite{ISDIL3D,JANKE} covariance
error-reduction was presented for the Finite-Size Scaling analysis of
phase transitions. Indeed, covariance analysis improves the
computation of the critical temperature and the leading
scaling-corrections exponent from data on finite lattices
~\cite{ISDIL3D}. It provides as well the optimal combination of
different estimates of the sought critical exponent (each individual
estimate being previously extrapolated to infinite
volume)~\cite{JANKE}. As we discuss in  Sect.~\ref{SECT-VARIAS-As},
covariance error reduction (specially as presented in Reference~\cite{JANKE}) is
a particular case of the present approach.

The layout of the rest of this note is as follows. In
Sect.~\ref{COV-SECT} we recall the error reduction strategy in a
general setting (without reference to any specific model). The reader
merely interested in a practical recipe, may proceed directly to
Sect.~\ref{SECT-PRACTICAL}. In Sect.~\ref{MODEL-SECT}, we briefly
describe the model and the observables, as well as the used mean-value
identities. We present our numerical results in
Sect.~\ref{RESULTS-SECT} while our conclusions are in
Sect.~\ref{CONCLUSIONS-SECT}. In Appendix~\ref{APENDICE-CLUSTER} we
present some technical results which are specific for the Swendsen-Wang cluster
algorithm as applied to the Ising model.

\section{Covariance error reduction}\label{COV-SECT}

We first discuss the problem as if the exact covariance matrix was
accessible (Sect.~\ref{SECT-MIN-ERR}). The effects of time
correlations are described in Sect.~\ref{SECT-TIME}. Real life
complications arise from the fact that the covariance matrix needs to
be estimated from a finite sample of Monte Carlo data, which
fortunately does not induce any significant bias,
Sect.~\ref{SECT-PRACTICAL}. Finally, we discuss in
Sect.~\ref{SECT-VARIAS-As} how the general approach relates with the
problem of finding the optimal linear combination of several estimates
for the very same expectation value. We discuss as well some of the
very counterintuitive features of this problem.

\subsection{The minimal  error} \label{SECT-MIN-ERR}
Let $A$, $B_1$, $B_2$,\ldots $B_R$ be stochastic variables. We assume
that a set of mean-value identities appropriate for the problem at hand
tell us that $\langle B_i\rangle=0$ for $i=1,2,\ldots R$.  We assume
as well that $\langle A^2\rangle$ and all the $\langle B_i^2\rangle$ are
finite.  We wish to profit from the covariance between $A$ and the
$B_i$ to obtain the best determination (in the sense of minimal
variance) of $\langle A\rangle$.

Before going on, it is useful to note that the operation of computing
the covariance between real-valued stochastic variables $X$ and $Y$,
\begin{equation}
\sigma_{XY}\equiv\langle
(X-\langle X\rangle) (Y-\langle Y\rangle)\rangle\,,
\end{equation} 
has the structure of a scalar product. Indeed the four following
properties are easy to establish (i) it is symmetric,
$\sigma_{XY}=\sigma_{YX}$, (ii) it is linear on each of its arguments,
$\sigma_{X (\lambda_1 Y_1 +\lambda_2 Y_2)}=\lambda_1
\sigma_{XY_1}+\lambda_2 \sigma_{XY_2}$, (iii) $\sigma_{XX}\geq 0$, and
(iv) if $\langle X \rangle=0$ and $\sigma_{XX}=0$ it follows that
$X=0$ with probability one. For later use, we introduce the
correlation coefficient between $X$ and $Y$
\begin{equation}
r_{XY}\equiv \frac{\sigma_{XY}}{\sqrt{\sigma_{XX} \sigma_{YY}}}\label{DEF-CORR-COEF}
\end{equation}

Using the $B_i$, it is straightforward to define stochastic variables
with expectation value $\langle A\rangle $:
\begin{equation}
\tilde A(\lambda_1,\lambda_2,\ldots,\lambda_R)= 
A +\sum_{i=1}^R \lambda_i B_i\,.\label{TILDEA-DEF}
\end{equation}
Our task is to find the coefficients $\{\lambda_i^*\}_{i=1}^R$ that
minimize the $\tilde A$ variance
\begin{equation}
\sigma_{\tilde A \tilde A}= \sigma_{AA} + 2\sum_{i=1}^R \lambda_i \sigma_{AB_i} +
\sum_{i=1}^R\lambda_i^2 \sigma_{B_i B_i}\, ,
\end{equation}
that has a minimum at
\begin{equation}
\lambda^*_i=-\sum_{i,i'=1}^R(\varSigma^{-1})_{i,i'} \sigma_{AB_{i'}},\quad \varSigma_{ii'}=\sigma_{B_iB_{i'}}\label{LAMBDA-OPT}
\end{equation}

In the following, we will
denote the optimal random variable as 
\begin{equation}
A^*=\tilde A(\lambda_1^*,\ldots,\lambda_2^*)\,\label{A*}
\end{equation}
whose variance is
\begin{equation}
\sigma_{A^{\!*}\!A^{\!*}}=\sigma_{AA}-\sum_{i,i'=1}^R \sigma_{AB_i} (\varSigma^{-1})_{ii'}\sigma_{AB_{i'}}\label{SIGMA-A*}
\end{equation}
Note that rescaling any of the $B_i$, $B_i\rightarrow \alpha_i B_i$,
would leave $A^*$ unchanged. For $R=1$, Eq.~(\ref{SIGMA-A*}) reads
\begin{equation}
\sigma_{A^{\!*}\!A^{\!*}}=\sigma_{AA}(1-r_{AB}^2)\;.\label{SIGMA-A*-R=1}
\end{equation}
In particular, whether $A$ and $B$ are correlated or anticorrelated is immaterial.

In a nutshell, we face a standard problem of best approximation in an
Euclidean space: we are decomposing the fluctuating part of $A$,
$A-\langle A\rangle$, on its components parallel and orthogonal with
respect to the linear space generated by $\{ B_i\}_{i=1}^R$. The best
approximation, $A^*$, is found when the parallel component is made to
vanish. The minimal variance is the norm squared of the orthogonal component.
If we compute in a Monte Carlo simulation $A^*$ rather than $A$, we
are rewarded with a CPU gain factor of $\sigma_{AA}/\sigma_{A^{\!*}\!A^{\!*}}$.

\subsection{Covariance and time correlations}\label{SECT-TIME}

The stochastic variables $X,Y,Z,\ldots$ considered in
Sect.~\ref{SECT-PRACTICAL} are actually Monte Carlo time averages.

Indeed, the Monte Carlo dynamics can be regarded as a Markovian
random-walk in configuration space~\cite{SOKALLECTURE}. Let $\varTheta$ be 
one of such spin (or gauge-field) configurations, and
$\varTheta_{t=0}$, $\varTheta_{t=1},\ldots\,$ be the time sequence of
configurations visited by the random-walker. We consider functions of
the fields configuration ${\cal X},{\cal Y}, {\cal Z},\ldots$
(observables, hereafter), and use the shorthand ${\cal X}^{(t)}={\cal
  X}[\varTheta(t)]$, $t=0,1,\ldots,T-1$. Hence, our stochastic variable $X$
will be (and similarly for $Y$, $Z$,\ldots)
\begin{equation}
X=\frac{1}{T} \sum_{t=0}^{T-1} {\cal X}^{(t)}\,.\label{X-IT-DEF}
\end{equation}

The {\em Markovian\/} random-walk in configuration space, is fully
determined by a transition matrix, ${\cal P}_{\varTheta_{t+1}\varTheta_t}$,
namely the conditional probability of reaching $\varTheta_{t+1}$ from
$\varTheta_t$ in a single step. The transition matrix verifies the {\em
  balance\/} condition, with respect to the equilibrium distribution
function $\pi[\varTheta]$
\begin{equation}
\pi[\varTheta_{t+1}]=\sum_{\varTheta_t}  {\cal P}_{\varTheta_{t+1}\varTheta_t} \pi[\varTheta_t]\,.\label{BALANCE-DEF}
\end{equation}
In this work, we shall always consider that, at $t=0$, equilibrium has
been already reached.  Thus, the expectation value for $X$ is the
Boltzmann average for ${\cal X}\,$. 

It is convenient to consider the equilibrium (symmetrized) time
correlation function for two real observables, ${\cal X}$ and ${\cal Y}$ 
({\em autocorrelation\/} if ${\cal X}={\cal Y}$)
\begin{equation}
C_{{\cal X}{\cal Y}}(t)=\frac12\langle {\cal X}^{(0)}{\cal Y}^{(t)}+ {\cal X}^{(t)}{\cal Y}^{(0)}\rangle- \langle {\cal X}\rangle \langle {\cal Y}\rangle\,.
\label{CXY-DEF}
\end{equation}
Note that $C_{{\cal X}{\cal Y}}(t)\!=\!C_{{\cal Y}{\cal
    X}}(t)\!=\!C_{{\cal X}{\cal Y}}(-t)$, and that it is bilinear in
${\cal X}$ and ${\cal Y}$. We will call $C_{{\cal X}{\cal Y}}(0)$ 
{\em static covariance\/},
since it can be computed from equal-time
expectation values. $C_{{\cal X}{\cal Y}}(t)$ allows to compute
$\sigma_{XY}$, since one straightforwardly obtains from \eqref{X-IT-DEF} that
\begin{equation}
\sigma_{XY}=\frac{1}{T^2} \sum_{t,t'=0}^{T-1}\ C_{{\cal X}{\cal Y}}(t'-t)\;.
\end{equation}

We define the integrated correlation time ({\em autocorrelation\/}
time, $\tau_{\text{int},{\cal X}}$, if ${\cal X}={\cal Y}$) as
\begin{equation}
\tau_{\text{int},{\cal X}{\cal Y}} =\frac{\sum_{t=-\infty}^{t=\infty}\, C_{{\cal X}{\cal Y}}(t)}
{2\,\sqrt{C_{{\cal X}{\cal X}}(0)C_{{\cal Y}{\cal Y}}(0)}}\,.\label{TAU-INT-DEF}
\end{equation}
Now, a standard argument~\cite{SOKALLECTURE} tells us that, if $\sum_{t=1}^\infty\ t\, |C_{{\cal X}{\cal Y}}(t)|<\infty$, the covariance of $X$ and $Y$ is
\begin{equation}
\sigma_{XY}=\frac{ 2\tau_{\text{int},{\cal X}{\cal Y}}\sqrt{C_{{\cal X}{\cal X}}(0)C_{{\cal Y}{\cal Y}}(0)}}{T} +{\cal O} (T^{-2})\,.
\end{equation}
For instance, the $r_{AB}$ in Eq.(\ref{SIGMA-A*-R=1}) is just
\begin{equation}
r_{AB}=\frac{\tau_{\text{int},{\cal A}{\cal B}}}{\sqrt{\tau_{\text{int},{\cal A}}\,
\tau_{\text{int},{\cal B}}}}\,.
\end{equation}
Hence, the effectivenes of a particular control variate, $B$, does
depend on the autocorrelation and correlation times of the chosen
Monte Carlo algorithm\footnote{For disordered systems, the time
  average in Eq.(\ref{X-IT-DEF}) is followed by a disorder-average,
  which strongly diminishes the influence of the particular Monte
  Carlo dynamics.}.

We finally recall some well known results~\cite{SOKALLECTURE}.
$C_{{\cal X}{\cal Y}}(t)$ can be computed from the $t$-th power of the
transition matrix and the equilibrium distribution as
\begin{eqnarray}
C_{{\cal X}{\cal Y}}(t)&=&\sum_{\varTheta_t,\varTheta_0}\ \frac{1}{2} 
\left({\cal X}[\varTheta_0] {\cal Y}[\varTheta_t] + {\cal X}[\varTheta_t] {\cal Y}[\varTheta_0] \right)\\\nonumber
&&\times\left([{\cal P}]^{|t|}_{\varTheta_t\varTheta_0}  - \pi[\varTheta_t]\right) \pi[\varTheta_0]\,.
\end{eqnarray}
At this point, an analogy with Quantum Mechanics is in order. Up to
now, we have been working in the Schr\"odinger picture, where the
probabilities evolve in time while the operators remain constant. Yet,
it is best to work in an equivalent Heisenberg picture where only
observables evolve in time. We define a time-transformation, $P$, that
transforms observable ${\cal X}$ in observable $P{\cal X}$. The value
taken by $P{\cal X}$ for configuration $\varTheta$ is a conditional
expectation value
\begin{equation}
P{\cal X}[\varTheta]=E({\cal X}[\varTheta_{t+1}] | \varTheta_{t}=\varTheta) =
\sum_{\varTheta'}{\cal X}[\varTheta'] {\cal P}_{\varTheta'\varTheta}\,.
\end{equation}
Mind that, if the Monte Carlo dynamics is composed of consecutive
steps (in the Swendsen-Wang dynamics, for instance, one first update
the bonds, then the spins: ${\cal P}^\text{SW}= {\cal P}_\text{spin}
{\cal P}_\text{bond}$), the evolution operators in the Heisenberg
picture appear in reversed order (e.g.  $P^\text{SW}=
P_\text{bond}P_\text{spin}$). We introduce a scalar product for {\em
  equal time\/} real observables $({\cal X},{\cal Y})\equiv \langle{\cal
  X}(t){\cal Y}(t) \rangle$. The correlation function is
\begin{equation}
C_{{\cal X}{\cal Y}}(t)=\frac{({\cal X}, P^{|t|} {\cal Y})+
(P^{|t|}{\cal X},  {\cal Y})}{2} 
-\langle{\cal X} \rangle \langle{\cal Y} \rangle\,.\label{C-HEISENBERG}
\end{equation}
Thus the problem of computing correlation times is reduced to the
spectral analysis of the operator $P$.

\subsection{Practical recipes}\label{SECT-PRACTICAL}

In a Monte Carlo calculation, the stochastic variables $A$ and $B_i$
discussed in Sect.~\ref{SECT-MIN-ERR} are directly related to some
functions of the spin (or gauge field) configuration, ${\cal A}$,
${\cal B}_i$, $i=1,2,\ldots,R$. One stores in disk $T$ consecutive
measurements of these functions $\{{\cal A}^{(t)},{\cal
  B}_1^{(t)},\ldots,{\cal B}_R^{(t)} \}_{t=1}^T$.  We assume that
autocorrelation times~\eqref{TAU-INT-DEF} for these measurements are
finite. Their Monte Carlo average
\begin{equation}
\overline A = \frac{1}{T}\sum_t {\cal A}^{(t)}\ ,\quad \overline B_i = \frac{1}{T}\sum_t {\cal B}^{(t)}_i\,, i=1,2,\ldots,R\,.\label{DEF-OVERLINE}
\end{equation}
are just instances (i.e. disorder realizations) of the random variables $A$ and $B_i$.  

Let us form $N$ {\em data blocks} $\{A_j,B_{i,j}\}_{j=1}^N$ by
averaging sets of $T/N$ consecutive measurements $\{{\cal
  A}^{(t)},{\cal B}_1^{(t)},\ldots,{\cal B}_R^{(t)} \}\,.$ The basic
assumption underlying the Monte Carlo error analysis~\cite{VICTORAMIT}
is that, provided that $T/N$ is large enough as compared to Monte
Carlo autocorrelation times, the $\{A_j,B_{i,j}\}_{j=1}^N$ are
identically distributed, and statistically independent for different
$j$. Furthermore, one assumes that  $T/N$
is so large, that the blocked data are not only independent, but also
Gaussian distributed:
\begin{equation}
\begin{array}{lcl}
  A_j&=&\displaystyle \langle A\rangle\ +\ \eta^A_j\,\sqrt{N \sigma_{AA}}\;,\\
\\
  B_{i,j}&=&\eta^{B_i}_j\,\sqrt{N \sigma_{B_i B_i}}\,,\ i=1,2,\ldots,R\;.\label{GAUSS1}
\end{array}
\end{equation}
 
The $\eta$ are Gaussian random numbers, with zero mean and
covariance matrix
\begin{eqnarray}
\langle \eta^A_j\eta^A_{j'}\rangle &=&\delta_{jj'}\;,\nonumber\\
\langle \eta^A_j\eta^{B_i}_{j'}\rangle&=& \delta_{jj'} r_{AB_i}\;,\label{GAUSS2}\\
\langle\eta^{B_i}_j\eta^{B_{i'}}_{j'}\rangle&=& \delta_{jj'} r_{B_i B_{i'}}\,,\nonumber
\end{eqnarray}
where $\delta_{jj'}$ is Kronecker's delta. Note as well that one
gets exactly the same numbers for $\overline A$ and $\overline B_i$ either by
averaging over $j$ the $\{A_j,B_{i,j}\}$, or using Eq.~(\ref{DEF-OVERLINE}).
For later use, we define also the jackknife blocks (see
e.g.~\cite{VICTORAMIT})
\begin{equation}
\begin{array}{lcl}
A^\mathrm{JK}_j&=&\displaystyle\frac{N \overline{A}-A_j}{N-1}\;,\\
\\
B^\mathrm{JK}_{i,j}&=&\displaystyle\frac{N \overline{B_i}-B_{i,j}}{N-1}\,, i=1,2,\ldots,R\,.
\end{array}
\end{equation}

Our statistical estimators for the covariances are
\begin{equation}
\begin{array}{lcl}
\overline{\sigma_{AA}}&=& \displaystyle\sum_{j=1}^N \frac{(A_j-\overline A)^2}{N(N-1)}\\
&=&\displaystyle\sum_{j=1}^N \frac{(A_j^\mathrm{JK}-\overline A)^2}{N/(N-1)}\,,\\\\
\overline{\sigma_{AB_i}}&=&\displaystyle\sum_{j=1}^N \frac{(A_j-\overline
  A)(B_{i,j}-\overline B_i)}{N(N-1)}\\
&=&\displaystyle\sum_{j=1}^N \frac{(A_j^\mathrm{JK}-\overline A)(B_{i,j}^\mathrm{JK}-\overline B_i)}{N/(N-1)}\,,\\\\
\overline{\sigma_{B_iB_{i'}}}&=&\displaystyle\sum_{j=1}^N \frac{(B_{i,j}-\overline B_i) (B_{i',j}-\overline B_{i'})}{N(N-1)}\\
&=&\displaystyle\sum_{j=1}^N \frac{(B_{i,j}^\mathrm{JK}-\overline B_i)(B_{i',j}^\mathrm{JK}-\overline B_{i'})}{N/(N-1)}\,.
\label{EST}\\
\end{array}
\end{equation}
At variance with the {\em numbers} $\sigma_{AA}$, $\sigma_{AB_i}$ or
$\sigma_{B_iB_j}$, our estimators $\overline{\sigma_{AA}}$,
$\overline{\sigma_{AB_i}}$ or $\overline{\sigma_{B_i B_{i'}}}$, are {\em
  random variables}. It is straightforward to show that their
expectation values are the sought covariances, but they are subject to
statistical errors whose (relative) size is of order $1/\sqrt{N}$. In
fact, since one needs to keep the data-block size $T/N$ as large as
possible to ensure the correctness of Eq.~(\ref{GAUSS1}), the typical
number of blocks is kept low, say $N\sim 100\,$.  Incidentally, the
second equality in each one of Eqs.~(\ref{EST})
is an algebraic one: we get the same numerical covariance estimates
from the standard or the jackknife blocks.

At this point, we may trade the unaccessible minimization Eqs.~(\ref{LAMBDA-OPT},\ref{A*})
by the computable
\begin{equation}
A^*= A -\sum_{i,i'=1}^R (\overline\varSigma^{\,-1})_{ii'}\overline{\sigma_{A
    B_{i'}}}\,B_i\ ,\quad \overline\varSigma_{ii'}=\overline{\sigma_{B_iB_{i'}}}\ .\label{A*-REAL}
\end{equation}
The very same procedure is performed block by block, thus obtaining $\{
A^*_j\}_{j=1}^N$. Errors are computed in a standard way from these blocks. 

The reader might question the validity of Eq.~(\ref{A*-REAL}), because
the vanishing of $\langle B_i\rangle$ does not imply $\langle
\sum_{i'} (\overline\varSigma^{\,-1})_{ii'}\overline{\sigma_{A
    B_{i'}}}\,B_i\rangle=0\,.$ This is specially worrying since, as we
said above, the relative errors for $\overline{\sigma_{AB_i}}$ or
$\overline{\sigma_{B_iB_{i'}}}$ are $\sim 10\%$ in real-life
calculations. The way-out is in Eqs.~(\ref{GAUSS1},\ref{GAUSS2}).  If
in a particular simulation one finds the Gaussian fluctuations
$\{\eta^A_j,\eta_j^{B_1},\ldots,\eta_j^{B_1}\}_{j=1}^N$, the
sign-reversed fluctuations
$\{-\eta^A_j,-\eta_j^{B_1},\ldots,-\eta_j^{B_1}\}_{j=1}^N$ are just as
probable. One immediately notices that the covariance estimators,
Eqs.~(\ref{EST}), are invariant under sign-reversal of
fluctuations. This means that $\overline\sigma_{A B_i}$, the matrix
$\overline\Sigma$ and its inverse are also invariant, while the $B_i$
transform to $-B_i$. Hence, if the probability distribution function
of $\{\eta^A_j,\eta_j^{B_1},\ldots,\eta_j^{B_1}\}_{j=1}^N$ is
invariant under sign-reversal, it follows that the expectation value
for $A^*$ in Eq.~(\ref{A*-REAL}) is still $\langle A\rangle$
(according to Rubinstein~\cite{RUBINSTEIN}, this fact was first
noticed for the particular case of Gaussian distributed fluctuations
in~\cite{IBM}). However, even in the absence of sign-reversal
invariance, the bias induced is of order $1/T$ while the statistical
error is of order $1/\sqrt{T}$.

As for functions of expectation values, let us explain the procedure by
considering the second moment correlation length Eq.~(\ref{XI-DEF}), that
depends on the expectation values of two variables, $m(0)$ and $m(\vec
k_\mathrm{min})$. One first transforms using Eq.~(\ref{A*-REAL}) the estimates
and the jackknife blocks of each of the needed quantities,
e.g. $\overline{m^*(0)}$, $\overline{m^*(\vec k_\mathrm{min})}$ and
$\{m_j^{\mathrm{JK},*}(0),m_j^{\mathrm{JK},*}(\vec k_\mathrm{min})\}_{j=1}^N$.
Then we use Eq.~(\ref{XI-DEF}) to obtain our best estimate of the correlation
length from $\overline{m^*(0)}$, $\overline{m^*(\vec k_\mathrm{min})}$. To
estimate the errors, we first form $N$ jackknife blocks by computing the
correlation length from each of the $N$ pairs
$\{m_j^{\mathrm{JK},*}(0),m_j^{\mathrm{JK},*}(\vec k_\mathrm{min})\}$, then use
the standard formulae~\cite{VICTORAMIT}.

\subsection{Several observables with same expectation value}\label{SECT-VARIAS-As}

Given a set of random variables $A_1,A_2,\ldots A_{R+1}$ with a common
expectation value, $\langle A_i\rangle=a$, one may wonder how to get the best
possible estimate of $a$. This was precisely the case considered
in~\cite{ISDIL3D,JANKE}. We only discuss here the relationship with the
(closer in spirit) approach
of~\cite{JANKE}, where the $A_i$ were estimates of the critical exponent 
$\nu$ for
an Ising model at its critical point. The obvious way of addressing the
problem is considering a linear combination
\begin{equation}
\tilde{A}(p_1,p_2,\ldots,p_{R+1})= \sum_{i=1}^{R+1} p_i A_i\;,\quad
\sum_{i=1}^{R+1} p_i =1\,,
\end{equation}
then minimizing $\sigma_{{\tilde A}{\tilde A}}$. This is a
particular case of the optimization problem that we have already
discussed at length in Sects.~\ref{SECT-MIN-ERR}
and~\ref{SECT-PRACTICAL}. In fact, note that
$p_{R+1}=1-p_1-p_2-\ldots-p_R$ and then, keeping an eye on
Eq.~(\ref{TILDEA-DEF}), write $A\equiv A_{R+1}$, $\{\lambda_i=p_i\,,\, B_i=A_i-A_{R+1}
\}_{i=1}^R$.

However, this optimization problem produced some counterintuitive
results~\cite{JANKE}.  All five computed $\nu$ estimates for the
two-dimensional Ising model lied {\em above} the exact value. In spite
of this, the improved estimate was {\em below} the exact value. This
apparent paradox can be easily explained in our language, by
considering the simpler case $R=1$, (so we have $A_1$ and
$A_2$). Using the results reviewed in Sect.~\ref{SECT-MIN-ERR} one easily finds that the
minimal squared error is
\begin{equation}
\sigma_{A^{\!*}\!A^{\!*}}=\frac{\sigma_{A_1A_1}\sigma_{A_2A_2}(1-r_{A_1A_2}^2)}
{\sigma_{A_1A_1}+\sigma_{A_2A_2}-2r_{A_1A_2}\sqrt{\sigma_{A_1A_1}\sigma_{A_2A_2}}}\,,
\end{equation}
Hence, if $r_{A_1A_2}$ tends to one {\em and} if $\sigma_{A_1A_1}\neq
\sigma_{A_2A_2}$ an error-less estimator exists. In fact, in the
$r_{A_1A_2}\to 1$ limit we have $A_1=a +\eta \sqrt{\sigma_{A_1A_1}}$,
$A_2=a +\eta \sqrt{\sigma_{A_2A_2}}$ with $\eta$ the {\em same} Gaussian
random-number for both variables (of course $\langle\eta\rangle=0$ and
$\langle\eta^2\rangle=1$). In other words, if for a particular
simulation $A_1$ lies below(above) $a$, the same will be true for
$A_2$. In spite of this, if we write $p A_1+(1-p)A_2 = a +\eta
[\sqrt{\sigma_{A_2A_2}}+p(\sqrt{\sigma_{A_2A_2}}-\sqrt{\sigma_{A_1A_1}})]$
and set
$p=\sqrt{\sigma_{A_2A_2}}/(\sqrt{\sigma_{A_1A_1}}-\sqrt{\sigma_{A_2A_2}})$,
an exact answer is found. Note, however, that the problem becomes ill
conditioned when $\sigma_{A_1A_1}$ approaches $\sigma_{A_2A_2}$. In fact, if
the two variance coincide we gain nothing by considering $A_2$ in
addition to $A_1$, since in this case one would have $A_1=A_2$ with
probability one.

\section{Model, observables, mean-value identities}\label{MODEL-SECT}

We shall put to work the strategy in Sect.\ref{COV-SECT}, in the
standard benchmark, the Ising model in two
dimensions, for which many exact results exist, including exact
computations of some quantities in {\em
  finite-systems}~\cite{FERDINAND-FISHER} that can be directly
confronted with the Monte Carlo simulation.

The spins ${\cal S}_{\vec x}$ are placed in the nodes of a square lattice of side $L$
with periodic boundary conditions. The interaction
is restricted to lattice nearest neighbors, the partition function being
($\sum_{\{{\cal S}_{\vec x}\}}$: summation over the $2^{L^2}$ spin configurations):
\begin{equation}
Z=\sum_{\{{\cal S}_{\vec x}\}} \mathrm{exp}\left[\kappa \sum_{||\vec x-\vec
    y||=1} {\cal S}_{\vec x} {\cal S}_{\vec y}\right]\,,
\end{equation}
The system undergoes a second order phase transition at
$\kappa_\mathrm{c}=\log(1+\sqrt{2})/2$.

The main functions of the spins that we are considering are the energy,
and the Fourier transform of the spin field at zero and minimal momenta
($\vec k=(0,0)$ or $\vec k_\mathrm{min}=(2\pi/L,0)$)
\begin{equation}
e=\frac{1}{L^2} \sum_{||\vec x-\vec y||=1} {\cal S}_{\vec x} {\cal S}_{\vec y}\;,\quad 
m(\vec k)=\frac{1}{L^2} \sum_{\vec x} {\cal S}_{\vec x}\mathrm{e}^{\mathrm{i} \vec k\cdot \vec x}\,.\label{E-DEF}
\end{equation}

From $m(\vec k)$ we define the magnetic susceptibility
\begin{equation}
\chi= L^2\langle [m(0)]^2\rangle\,,\label{CHI-DEF}
\end{equation}
the second moment correlation length~\cite{COOPER} (we gain statistics
by averaging  $[m(\vec k_\mathrm{min})]^2$ over $(2\pi/L,0)$ and $(0,2\pi/L)$)
\begin{equation}
\xi=\sqrt{
\frac{\langle [m(0)]^2\rangle- \langle [m(\vec k_\mathrm{min})]^2\rangle}{4\, \mathrm{sin}^2\frac{\pi}{L}\, \langle [m(\vec k_\mathrm{min})]^2\rangle}}\,,\label{XI-DEF}
\end{equation}
and the Renormalization-Group invariant ratio
\begin{equation}
U_4=\frac{\langle [m(0)]^4\rangle}{\langle [m(0)]^2\rangle^2}\,.\label{U4-DEF}
\end{equation}

Our first mean-value identity comes from the Fortuin-Kasteleyn formulation
(see e.g.~\cite{SOKALLECTURE,VICTORAMIT} for details). Given a
decomposition of the lattice in ${\cal N}$ connected components ({\em
  clusters}), each containing $n_c$ spins, it is easy to show that
(see Appendix~\ref{APENDICE-CLUSTER} for a quick review)
\begin{equation}
\chi= \frac{1}{L^2}\langle \sum_{c} n_c^2 \rangle\,.\label{chi-CLUSTER}
\end{equation}  
Hence, our first control variate is
\begin{equation}
{\cal B}_\text{SW}= [m(0)]^2 - \sum_{c} \frac{n_c^2}{L^4}\,,\label{Bc-DEF}
\end{equation}

A second control variate comes from a Callen identity~\cite{CALLEN}. Let
the local field acting over site $\vec x$ be
\begin{equation}
h_{\vec x}=\sum _{||\vec x-\vec y||=1} {\cal S}_{\vec y}\,.
\end{equation}
Then, if $||\vec x-\vec y||>1$,
\begin{equation}
\langle {\cal S}_{\vec x} {\cal S}_{\vec y} \rangle=\langle \mathrm{tanh} (\kappa h_{\vec x})\,  \mathrm{tanh} (\kappa h_{\vec y})\rangle\,.
\end{equation}
Hence,
\begin{equation}
{\cal B}_\text{CI}=\frac{1}{L^4}\sum_{\Vert\vec x-\vec y\Vert>1}[\mathrm{tanh}(\kappa h_{\vec
  x})\mathrm{tanh}(\kappa h_{\vec y})\ -{\cal S}_x {\cal S}_y]\,,\label{Bs-DEF}
\end{equation}
that can be computed with ${\cal O}(L^2)$ operations as
\begin{eqnarray}
{\cal B}_\text{CI}&=& \frac{1}{L^4}\left[\left(\sum _{\vec x} \mathrm{tanh} (\kappa
  h_{\vec x})\right)^2-\left(\sum_{\vec x} {\cal S}_{\vec x}\right)^2\right.\\
\nonumber\\\nonumber\\
&&\left.\qquad-\sum_{\vec x}[\mathrm{tanh}^2 (\kappa h_{\vec x})-1]\right.\nonumber\\
&&\qquad\left.-\sum _{\Vert\vec x-\vec y\Vert=1} [\mathrm{tanh} (\kappa
  h_{\vec x})\,  \mathrm{tanh} (\kappa h_{\vec y})\ -\ {\cal S}_{\vec x} {\cal
    S}_{\vec y}]\right]\,.\nonumber
\end{eqnarray}

Finally, a Schwinger-Dyson equation~\cite{SD} provides a third control variate
\begin{equation}
{\cal B}_\text{SD}= 1- \frac{1}{L^2} \sum_{\vec x} \mathrm{e}^{-2\kappa h_{\vec x}}\,.
\end{equation}

\section{Results}\label{RESULTS-SECT}

\begin{table*}
\begin{center}
\begin{tabular*}{\textwidth}{@{\extracolsep{\fill}}lllll}
\hline
$L$ & $\langle e\rangle$ & $\chi$ &  $\xi$ & $U_4$ \\
\hline
\hline
{\bf 16} &&&& \\
standard & 1.45339(\textbf{47}) & 139.719(\textbf{\textbf{155}})& 14.601(\textbf{31}) &1.16502(\textbf{74}) \\
cluster  &             & 139.713(\textbf{127})&   &   \\
$B_\text{SW}$ improved &1.45334(\textbf{32})&139.700(\textbf{93}) &14.597(\textbf{19})&1.16510(\textbf{51}) \\
$B_\text{CI}$ improved &1.45316(\textbf{24})&139.652(\textbf{104})&14.590(\textbf{25})&1.16524(\textbf{63})\\
$B_\text{SW}$ \& $B_\text{CI}$ improved  & 1.45319(\textbf{18}) & 139.666(\textbf{73})& 14.594(\textbf{18}) & 1.16517(\textbf{50})\\
others   & 1.453065\ldots\cite{FERDINAND-FISHER}&
139.546(\textbf{77})~\cite{TETHERED}& 14.566(\textbf{14})~\cite{TETHERED}&
1.16586(\textbf{34})~\cite{SALAS-SOKAL}
\\\hline
{\bf 128} &&&& \\
standard &1.419052(\textbf{100})& 5316.6(\textbf{76})&115.77(\textbf{28})& 1.16789(\textbf{89})\\
cluster  &           & 5317.7(\textbf{70})&   &   \\
$B_\text{SW}$ improved &1.419101(\textbf{94})&5321.7(\textbf{60})&115.97(\textbf{21})&1.16735(\textbf{75})\\
$B_\text{CI}$ improved &1.419047(\textbf{79})&5316.4(\textbf{68})&115.77(\textbf{26})&1.16791(\textbf{86})\\
$B_\text{SW}$ \& $B_\text{CI}$ improved &1.419095(\textbf{66})&5321.4(\textbf{51})&115.96(\textbf{19})& 1.16736(\textbf{71})\\
others &1.419076\ldots\cite{FERDINAND-FISHER}&5318.1(\textbf{28})~\cite{TETHERED}&115.81(\textbf{13})~\cite{TETHERED}&1.16763(\textbf{32})~\cite{SALAS-SOKAL}\\\hline
{\bf 512} &&&& \\
standard &1.415407(\textbf{36})& 60180(\textbf{94})& 463.62(\textbf{115})& 1.16809(\textbf{89})\\
cluster  &            & 60168(\textbf{88})&&\\
$B_\text{SW}$ improved &1.415397(\textbf{34})&60134(\textbf{80})&462.99(\textbf{92})&1.16852(\textbf{76})\\
$B_\text{CI}$ improved &1.415429(\textbf{26})&60230(\textbf{78})&464.14(\textbf{101})&1.16768(\textbf{78})\\
$B_\text{SW}$ \& $B_\text{CI}$ improved &1.415421(\textbf{24})& 60183(\textbf{62}) & 463.51(\textbf{76}) & 1.16812(\textbf{64})\\
others &1.415429\ldots\cite{FERDINAND-FISHER}&60209(\textbf{34})~\cite{TETHERED}&
463.82(\textbf{51})~\cite{TETHERED}
& 1.16782(\textbf{30})~\cite{SALAS-SOKAL} \\\hline

\end{tabular*}
\end{center}
\caption{Comparison of numerical results for the quantities defined in
  Eqs.~(\ref{E-DEF},\ref{CHI-DEF},\ref{XI-DEF},\ref{U4-DEF}), namely
  the internal energy, the magnetic susceptibility, the correlation
  length and the dimensionless ratio $U_4$, as obtained in the two
  dimensional Ising model at its critical point, for different lattice
  sizes. For the susceptibility we show also the cluster estimate,
  Eq.~(\ref{chi-CLUSTER}), that improves less than a 20\% in terms of
  CPU time over the standard estimator Eq.~(\ref{CHI-DEF}).  In
  contrast, the covariance improved estimates obtained from the mean-value
  identities Eqs.~(\ref{Bc-DEF},\ref{Bs-DEF}) do save more than a
  factor 2 in computer cost. To check for the possibility of a bias
  induced by the covariance error-reduction, we compare also with
  exact results (for the internal energy) or with independent and
  longer Monte Carlo simulations.}
\label{TABLA-REAL}
\end{table*}

We have simulated the model on systems $L=16,128$ and $512$ using the
Swendsen-Wang algorithm~\cite{SWENDSEN-WANG}. For each lattice size, we traced
clusters $10^6$ times taking measurements each time that the clusters were
traced. We discarded the first $10\%$ of measurements for thermalization
(which, on the view of the autocorrelation times for this model and
algorithm~\cite{VICTORAMIT}, is extremely conservative), hence formed $N=100$
data-blocks of $9000$ measurements each (we expect to be well in the Gaussian
fluctuations regime). The jackknife error was used throughout for error
computations. The used programs were minor modifications of the sample
programs in~\cite{VICTORAMIT}.

As in section~\ref{SECT-PRACTICAL}, we name $B_i$
($i=\text{SW,CI,SD}$) the block average of consecutive Monte Carlo
measurements of ${\cal B}_i$.  The results of the analysis using
$B_\text{SW}$ and/or $B_\text{CI}$ as control variates are shown in
Table~\ref{TABLA-REAL}. We detect no bias when comparing with exact
results or with previously published (and more precise) computations.
When using the two control variates together, a CPU factor gain larger
than two is achieved for $\chi$, $\langle e\rangle$ and $\xi$, for all
values of $L$. This gain is largest for $L=16$ and deteriorates
somewhat in going to $L=128$, but then stabilizes and does not
significantly deteriorate further when going to $L=512$.  For
instance, for $L=512$ the CPU gain in the computation of the
susceptibility is a factor 2.3 when comparing with the standard spin
estimate [Eq.~(\ref{CHI-DEF})] or a factor 2.0 when comparing with the
cluster estimate [Eq.~(\ref{chi-CLUSTER})].

Rather smaller gains are obtained by using $B_\text{SW}$ and/or
$B_\text{CI}$ individually: for instance, in the $\chi$ computation
using $B_\text{SW}$ alone, the CPU gain factor is $2.8$ for $L=16$,
but it deteriorates to $1.56$ for $L=128$ and $1.42$ for $L=512$.  The
fact that we do significantly better by combining the two control
variates (rather than using only one of them) suggests that the
orthogonal component of $B_\text{CI}$ with respect to $B_\text{SW}$ is
sizeable (and that this component still strongly correlates with the
squared magnetization).

There are some interesting issues regarding the usefulness of
$B_\text{SW}$ as a control variate for $\chi$. This is an instance of
the problem considered in Sect.~\ref{SECT-VARIAS-As}: we are after the
optimal linear combination between
Eqs.~(\ref{CHI-DEF},\ref{chi-CLUSTER}).  In
Appendix~\ref{APENDICE-CLUSTER} we show that the optimal choice is
very close to the cluster-based susceptibility,
Eq.~(\ref{chi-CLUSTER}) (the optimum is exactly \eqref{chi-CLUSTER} if
successive measurements are separated by a time interval of many
autocorrelation times, so that they are essentially statistically
independent). This statement can be reworded as {\em the use of the
  spin-based susceptibility via the control variate $B_\text{SW}$
  barely improves the cluster-based susceptibility} (the usefulnes of
$B_\text{SW}$ decreases with growing autocorrelation times,
Eq.~\eqref{rCBSW-FINAL}).

A related, yet different, issue is the temperature evolution of the
efficency of the cluster estimator for $\chi$. At $\kappa_\mathrm{c}$,
Table~\ref{TABLA-REAL}, errors for the spin and cluster based
estimates are similar. This is in marked contrast with the situation
in the paramagnetic scaling region ($\kappa<\kappa_\mathrm{c}$,
$1\ll\xi\ll L$), see e.g.~\cite{TETHERED}. In
Eq.~\eqref{TILDE-R2-FINAL}, we give the (squared) ratio of statistical
errors for the two estimators in terms of an autocorrelation time and
of several expectation values of the {\em static} cluster-sizes
distribution. At $\kappa_\mathrm{c}$, a giant cluster dominates sums
such as that in~\eqref{chi-CLUSTER}, see Table~\ref{TABLA-VARIANCIAS}
in the Appendix. As a consequence, the squared error ratio at
$\kappa_\mathrm{c}$, Eq.~\eqref{TILDE-R2-FINAL}, is $\sim
1+\frac{1.15}{\tau_{\text{int},C}}$, never very large since
$\tau_{\text{int},C}\geq 1/2$, and decreasing with growing $L$ due to
critical slowing down. On the other hand, in the scaling region the
largest cluster is not dramatically large, and a major (static)
variance reduction is achieved by averaging over the sign of the
different clusters at a fixed time.  This gain is at the level of a
single measurement. Yet, Eq.~\eqref{TILDE-R2-FINAL}, the benefits 
remains after that the Monte Carlo time averaging.

As for the benefits of including $B_\text{SD}$ in the covariance reduction
procedure, they are marginal at the critical point (the CPU gained when adding $B_\text{SD}$ to  $\{B_\text{SW},B_\text{CI}\}$ is less than a 10\%).
Nevertheless, in the scaling region it can pay to consider
$B_\text{SD}$. For instance, in a $L=512$ lattice at $\kappa=0.42$, where
$\xi\sim 12$, we obtain a CPU gain factor of 1.23 for the cluster estimator of
the susceptibility, and 1.6 factor for the energy.

\section{Conclusions}\label{CONCLUSIONS-SECT}

For any problem amenable to a path-integral formulation there are well
known strategies (Schwinger-Dyson~\cite{SCHWINGER-DYSON},
Callen~\cite{CALLEN}, etc.) to obtain identities, that imply the
vanishing of a precise linear combination of expectation values of
non-trivial observables.  More often than not, researchers performing
Monte Carlo simulations compute the quantities appearing in the
identities, since the extra CPU costs is negligible and the identities
provide important consistency tests. In particular, they allow to
detect easily problems as frightening as programming bugs, failure of
the used pseudo random number generator, or thermalization bias. What
we have pointed out here is that, using the general and simple {\em
  control variates} strategy~\cite{HAMMERSLEY,RUBINSTEIN}, these identities
provide as well a significant error reduction in the final outcome of
Monte Carlo simulations. This comes at negligible CPU cost. The method has
been exemplified in the standard benchmark, the two-dimensional Ising
model at criticality.

We note nevertheless that less trivial applications of this technique already
exist. In particular, we have found that a Schwinger-Dyson equation providing
a now standard thermalization test in spin-glass simulations~\cite{YOUNG}, can
gain an error reduction factor of one half on some final quantities (e.g.
the correlation length)~\cite{HSGPT}.

\section*{Acknowledgments}
We acknowledge
partial financial support from Ministerio de Ciencia e Innovaci\'on 
(Spain) through research contract FIS2006-08533.

\appendix
\section{On cluster estimators}\label{APENDICE-CLUSTER}

We will answer here two related questions: (1) Why the control variate
$B_\text{SW}$ improves so little the {\em cluster\/} estimate of the
susceptibility [\eqref{chi-CLUSTER}] and (2) why, at the critical
point, the susceptibility cluster estimator barely improves over the
spin one [\eqref{CHI-DEF}].  Both questions are specific to the
Swendsen-Wang dynamics for ferromagnetic systems~\cite{EXPLICACION}.

Under a simplifying assumption, Question 1 is addressed in the
Sect.~\ref{BSW-SECT}, while Question 2 is considered in
Sect.~\ref{SECT-VARIANCIA-ESTATICA}.  The assumption is that
successive measurements are separated by a time interval of many
autocorrelation times, so that they are essentially statistically
independent. The assumption is removed in Sect.~\ref{SECT-LI-SOKAL}
(largely inspired in Ref.~\cite{LI-SOKAL}). Yet, the static variance
ratio computed in Sect.~\ref{SECT-VARIANCIA-ESTATICA} still plays a
prominant role in the general case.

\subsection{$\boldsymbol B_\text{SW}$ for independent measurements}\label{BSW-SECT}

For independent measurements, time correlation functions,
Eq.~\eqref{CXY-DEF}, vanish for all times $t\neq 0$. Hence, we need
only to compute a static covariance.

Let ${\cal M}\equiv L^2 m(0)$ be the extensive magnetization
(recall Sect.~\ref{MODEL-SECT}). At time $t$ in the
Swendsen-Wang dynamics, the lattice will be decomposed in ${\cal N}_t$
connected components, of size $n_c^t$ with $c=1,2,\ldots {\cal N}_t$
(the ordering is such that $n_1^t\geq n_2^t\geq n_3^t \ldots\,$). All
the spins belonging to cluster $c$ are given a common sign, ${\cal
  S}_c^t$. The value ${\cal S}_c^t=\pm 1$ is chosen with $50\%$
probability, independently for each cluster $c$~\cite{EXPLICACION}.

The spin-estimator for ${\cal M}^2$ is
\begin{equation}
{\cal M}^2_t = \sum_{c,c'=1}^{{\cal N}_t} n^t_c n^t_{c'} {\cal S}_c^t {\cal S}^t_{c'}\,.\label{EQ-M2-espin}
\end{equation}
On the other hand, if one averages Eq.~(\ref{EQ-M2-espin}) over the
$2^{{\cal N}_t}$ equivalent choices for the ${\cal S}^t_c=\pm 1$, only
the diagonal terms $c=c'$ survive.  Hence, the natural cluster
estimator for $\langle{\cal M}^2\rangle$ is the Monte Carlo average of
\begin{equation}
{\cal C}_t= \sum_{c=1}^{{\cal N}_t} (n^t_c)^2\,.
\end{equation}

It is illuminating to write Eq.~(\ref{EQ-M2-espin}) as
\begin{eqnarray}
{\cal M}^2_t &=& \langle {\cal M}^2\rangle + \eta_{\cal C}^t + 
\eta_{\cal S}^t\,,\label{M2-ASTUTA}\\
\eta_{\cal C}^t&=& {\cal C}_t - \langle {\cal M}^2\rangle\,,\\
\eta_{\cal S}^t&=& \sum_{c\neq c'} n^t_c n^t_{c'} {\cal S}_c^t {\cal
  S}^t_{c'}\   ,\quad {\cal B}_{\text{SW},t}=\frac{\eta_{\cal S}^t}{L^4}\,.
\end{eqnarray}
Of course, $\langle \eta_{\cal C}^t\rangle=\langle {\eta_{\cal
    S}^t}\rangle$=0, but the statistical independence of the ${\cal S}_c^t$
also implies $\langle \eta_{\cal C}^t \eta_{\cal S}^t\rangle =0$. 
Therefore,
\begin{equation}
\sigma_{{\cal C}{\cal C}}= \langle \eta_{\cal C}^2\rangle\ ,\quad 
\sigma_{{\cal M}^2{\cal M}^2}= \langle \eta_{\cal C}^2\rangle+\langle \eta_{\cal S}^2\rangle\,.
\end{equation}

Let us try
to improve ${\cal C}$ using ${\cal B}_\mathrm{SW}$ as control variate. We find
$C_{{\cal C}{\cal B}_\text{SW}}(0) = \langle \eta_{\cal C}^t \eta_{\cal
  S}^t\rangle/L^4=0$.  It follows that the improved estimator ${\cal C}^*$
obtained using ${\cal B}_\text{SW}$ as control variate is just ${\cal C}$. Using the
language of section~\ref{SECT-VARIAS-As}: with no time correlations, the optimal
linear combination between 
$L^2[m(0)]^2$ and $L^{-2}\sum_c n^2_c$ is just $L^{-2}\sum_c n^2_c$.

\subsection{The static variance}\label{SECT-VARIANCIA-ESTATICA}

\begin{table*}[t]
\begin{center}
\begin{tabular*}{\textwidth}{@{\extracolsep{\fill}}llllll}
\hline
$\kappa=\kappa_{\mathrm{c}}$, $y=(D+2-\eta)/2$&&&&&\\
\hline
$L$ & $g_{\cal C}$ & $R$ & $\langle n_1 \rangle L^{-y}$ 
& $\langle n_2/n_1\rangle$ & $\langle n_3/n_1\rangle$ \\
\hline
16 &0.11590(43)&1.19371(35)&1.00701(60)&0.12953(41)&0.06484(21)\\
128&0.12440(62)&1.16125(51)&1.00687(84)&0.12528(49)&0.06180(27)\\
512&0.12572(62)&1.15683(41)&1.00683(93)&0.12485(52)&0.06166(29)\\
\hline
\hline
$\kappa=0.42$, $y=0$&&&&&\\
\hline
16 &0.28942(73)&1.22644(53)&137.68(12)&0.27402(53)&0.15343(34)\\
128&0.16134(80)&3.4069(80)&1016.78(66)&0.72070(22)&0.58116(24)\\
512&0.010212(22)&13.967(15)&1963.27(60)&0.82843(14)&0.73909(15)\\
\hline
\end{tabular*}
\end{center}
\caption{Numerical determinations for different lattice sizes, both at
  $\kappa_\text{c}$ and at $\kappa=0.42$ (where $\xi\sim 12$ for large
  $L$), of the dimensionless ratios $g_{\cal C}$ and $g_{\cal S}$,
  Eq.~(\ref{DEFINE-Gs}), and the cluster-estimator's merit number $R$,
  Eq.~(\ref{DEFINE-R}), recall also \eqref{R2-DEF}. Note that at
  $\kappa=0.42$ the advantages of using a cluster estimator grows
  fastly with $L$, while it remains fairly modest at
  $\kappa_\text{c}$. We show as well the product $\langle n_1
  L^{-(D+2-\eta)/2}\rangle$ at $\kappa_\text{c}$, where $n_1$ is the
  largest cluster, $D=2$ and $\eta=1/4$ is the anomalous dimension.
  We see that $n_1^2$ scales as the full sum $\sum_c n_c^2$ (indeed
  $\langle n_1\rangle^2 < \langle n_1^2\rangle < \langle\sum_c
  n_c^2\rangle = L^D \chi\propto L^{D+2-\eta}\,$). On the contrary, at
  $\kappa=0.42$, $n_1$ grows only mildly with $L$. We can also compare
  for both $\kappa$ the average ratio of the sizes of the
  second-largest to largest cluster ($n_2/n_1$), and that of
  third-largest to largest ($n_3/n_1$). While at $\kappa_\mathrm{c}$
  there is a $L$-invariant hierarchical structure $n_1\sim 8 n_2 \sim
  16 n_3\ldots\,$, at $\kappa=0.42$ the largest cluster becomes a
  typical one with growing $L$.  }
\label{TABLA-VARIANCIAS}
\end{table*}

Under the independent measurements assumption, the (squared) error
ratio for the spin [\eqref{CHI-DEF}] and cluster [\eqref{chi-CLUSTER}]
susceptibility estimators equals the static variance ratio
\begin{equation}
R^2=\frac{C_{{\cal M}^2{\cal M}^2}(0)}{C_{{\cal C}{\cal C}}(0)}\,.\label{R2-DEF}
\end{equation}

To relate $R^2$
with the cluster size distribution, we start from
Eq.~\eqref{M2-ASTUTA} and a trivial relation between $C_{{\cal M}^2{\cal M}^2}(0)$
and the dimensionless ratio $U_4$, Eq.~\eqref{U4-DEF}:
\begin{equation}
\frac{C_{{\cal M}^2{\cal M}^2}(0)}{\langle 
{\cal M}^2 \rangle^2}= \frac{\langle {\cal M}^4 \rangle - \langle {\cal M}^2 \rangle^2}{\langle 
{\cal M}^2 \rangle^2} =U_4 -1\,.\label{INTERPRETACION-BINDER}
\end{equation}
 Hence, in the scaling region, where $U_4 \approx 3$, the spin
 estimator will be remarkably noisier than at
 $\kappa_{\mathrm{c}}$, see Table~\ref{TABLA-REAL}.

The covariance matrix for the $\eta_{\cal S},\eta_{\cal C}$ can be expressed
in terms of the $n_c$\footnote{ Combining $\langle {\eta_{\cal C}^t\eta_{\cal
      S}^t}\rangle=0$ with
  Eqs.~(\ref{M2-ASTUTA},\ref{INTERPRETACION-BINDER},\ref{ERROR-DE-CLUSTER},\ref{ERROR-DE-ESPIN}),
  we recover the well known cluster estimator for $U_4$:
$$
U_4=\frac{\left\langle 3(\sum_{c} n_c^2)^2 - 2\sum_c n_c^4\right\rangle}{\left\langle \sum_{c} n_c^2\right\rangle^2}\,.
$$}:
\begin{eqnarray}
\langle \eta_{\cal C}^{2}\rangle &=& \biggl\langle\biggl(\sum_c n_c^2\biggr)^2\biggr\rangle -
\biggl\langle \sum_c n_c^2 \biggr\rangle^2\,,\label{ERROR-DE-CLUSTER}\\
\langle \eta_{\cal S}^2\rangle &=& 2\biggl\langle \biggl(\sum_{c} n_c^2\biggr)^2 - \sum_c
n_c^4\biggr\rangle\ ,\label{ERROR-DE-ESPIN}
\end{eqnarray}
so that
\begin{equation}
R= \sqrt{1+\frac{\langle \eta_{\cal S}^2\rangle}{\langle \eta_{\cal
      C}^2\rangle}}\,.\label{DEFINE-R}
\end{equation}
Introducing the dimensionless ratios
\begin{equation}
  g_{\cal C}=\frac{\langle \eta_{\cal C}^2\rangle}{\langle {\cal M}^2\rangle}\quad,\quad
  g_{\cal S}=\frac{\langle \eta_{\cal S}^2\rangle}{\langle {\cal M}^2\rangle}\,,\label{DEFINE-Gs}
\end{equation}
we note that
\begin{equation}
U_4-1 = g_{\cal C}+g_{\cal S}\ ,\quad R=\sqrt{1+\frac{g_{\cal S}}{g_{\cal C}}}\,.\label{BALANCE-VARIANCIAS}
\end{equation}

Now, in the paramagnetic scaling region ($1\ll\xi\ll L$) the
thermodynamic limit of $g_{\cal S}$ is 2. Indeed, the two terms in the
difference $\langle \eta_{\cal S}^2\rangle =2 \left\langle
\left(\sum_c n_c^2\right)^2 -\sum_c n_c^4\right\rangle$ scale
differently: when $\xi\ll L$ the first grows as the system volume {\em
  squared}, while the second scales linearly with
volume\footnote{For finite $L$, $g_{\cal C}>0$ while the large-$L$
  limit for $U_4$ is 3.}. As a consequence, $R$ {\em diverges} if one
takes the large-$L$ limit at fixed $\kappa<\kappa_{\mathrm{c}}$. Since
 the susceptibility $\chi=\langle {\cal
  M}^2/L^2\rangle$ remains finite for large $L$, the error incurred
when estimating the susceptibility from a {\em single\/} measurement,
${\cal C}_t$, vanishes in the large-$L$ limit.

Quite on the contrary, Eq.~(\ref{BALANCE-VARIANCIAS}), considered precisely at
$\kappa_{\mathrm{c}}$, strongly suggests that both $g_{\cal C}$ and $g_{\cal
  S}$ have a finite, non vanishing, large-$L$ limit (and hence a finite
$R$). 

We display in Table~\ref{TABLA-VARIANCIAS} our results for $g_{\cal
  C}$, and $U_4$ both at the critical point and at $\kappa=0.42$,
where $\xi\approx 12$. Indeed $R(\kappa_\mathrm{c})\sim 1.15$ remains
bound.  As we show in Table~\ref{TABLA-VARIANCIAS}, the average ratios
$n_2/n_1$, $n_3/n_1$ at $\kappa_\mathrm{c}$ are surprisingly small and
size independent. In other words, the two sums in
(\ref{ERROR-DE-ESPIN}) are dominated by $n_1$, causing a massive
cancelation that diminish $g_{\cal S}$ as compared to $g_{\cal C}$.

\subsection{Monte Carlo time-correlations}\label{SECT-LI-SOKAL}

We now drop the assumption of independent measurements. The
(squared) ratio of the errors of the spin and susceptibility estimators
is no longer $R^2$, \eqref{R2-DEF}, but
\begin{equation}
\tilde R^2=\frac{\sum_{t=-\infty}^{t=\infty}\ C_{{\cal M}^2{\cal M}^2}(t)}
{\sum_{t=-\infty}^{t=\infty}\ C_{{\cal C}{\cal C}}(t)}
\end{equation}
Similarly, Eq.~\eqref{SIGMA-A*-R=1}, the efficiency of $B_\text{SW}$ as control variate to improve
the cluster susceptibility estimator is ruled by the correlation coeficient
\begin{equation}
r_{CB_\text{SW}}=\frac{\sum_{t=-\infty}^{t=\infty}\, C_{{\cal C}{\cal B}_\text{SW}}(t)}
{\left[\sum_{t=-\infty}^{t=\infty}\, C_{{\cal C}{\cal C}}(t) \right]^{\frac12}
\left[\sum_{t=-\infty}^{t=\infty}\, C_{{\cal B}_\text{SW}{\cal B}_\text{SW}}(t)\right]^{\frac12}}
\end{equation}
Arguing as in Ref.~\cite{LI-SOKAL} will lead us to our main result:
\begin{eqnarray}
\tilde R^2&=&1+\frac{1}{2\tau_{\text{int},{\cal C}}}\left[R^2+1\right]\,,\label{TILDE-R2-FINAL}\\
r_{CB_\text{SW}}&=& \frac{1}{\left[2\tau_{\text{int},{\cal C}}\, (R^2-1)\right]^{\frac12}}\,.\label{rCBSW-FINAL}
\end{eqnarray}

Since $R^2(\kappa_\mathrm{c})\sim 1.3$, the efficiency of the cluster
estimator at $\kappa_\mathrm{c}$ is ruled by $\tau_{\text{int},{\cal
    C}}$.  Indeed, the (mild) critical slowing down can
be traced in Table~\ref{TABLA-REAL}. The usefulness of
$B_\text{SW}$ as control variate, Eq.~\eqref{rCBSW-FINAL},
deteriorates as well with growing $\tau_{\text{int},{\cal C}}$.  

On the other hand, in the paramagnetic scaling region
($\kappa<\kappa_\mathrm{c}$, $1\ll\xi\ll L$) one easily has $\tilde
R^2\sim 100$ or larger. Given Eq.~\eqref{TILDE-R2-FINAL}, and since
$\tau_{\text{int},{\cal C}}\geq 1/2$ (because $C_{{\cal C}{\cal
    C}}(t)>0$, see below), this is due to the large $R^2$ that are to
be expected, recall Sect.~\ref{SECT-VARIANCIA-ESTATICA} (we expect
$\tau_{\text{int},{\cal C}}$ to be upper-bounded in the large-$L$
limit, for $\kappa<\kappa_\mathrm{c}$). However,
Eq.~\eqref{rCBSW-FINAL}, in the scaling region, $B_\text{SW}$ behaves
poorly as a control variate, since $\tau_{\text{int},{\cal C}}$ is
lower-bounded while $R^2$ diverges in the large-$L$ limit.

To derive Eqs.~(\ref{TILDE-R2-FINAL},\ref{rCBSW-FINAL}) we first
note that (in space dimension $D$)
\begin{eqnarray}
L^{2D}C_{{\cal C}{\cal B}_\text{SW}}(t)&=&C_{{\cal C}{\cal M}^2}(t)-C_{{\cal C}{\cal C}}(t)\,,\label{C_CBSW}\\
L^{4D}C_{{\cal B}_\text{SW}{\cal B}_\text{SW}}(t)&=&C_{{\cal M}^2{\cal M}^2}(t)+C_{{\cal C}{\cal C}}(t)\nonumber\\&&-2 C_{{\cal C}{\cal M}^2}(t)\,
\end{eqnarray}

Eq.~(\ref{C-HEISENBERG}) suggests that it will be fruitful  to recall
the main properties of the operator
$P^\text{SW}=P_\text{bond}P_\text{spin}$. The two operators
$P_\text{bond}$ and $P_\text{spin}$ are of heat-bath type, and their
action is quite simple~\cite{EXPLICACION}: for any observable $O$,
$P_\text{spin}O=E(O|\{b\})$ and $P_\text{bond}O=E(O|\{{\cal S}\})$.
In particular, we have 
\begin{equation}
P_\text{spin}{\cal M}^2={\cal C}\,,\ 
P_\text{spin}{\cal C}={\cal C}\,,\ P_\text{bond}{\cal M}^2={\cal M}^2\,.
\end{equation}

All heat-bath operators, $P^\text{HB}$, share some nice features: they
are self-adjoint, $(O_1,P^\text{HB}O_2)\!=\!(P^\text{HB}O_1,O_2)$ and
idempotent $[P^\text{HB}]^2\!=\!P^\text{HB}$. Furthermore, they preserve
expectation values $\langle O\rangle\!=\!\langle P^\text{HB}O\rangle\,$
\footnote{ For instance, $\langle {\cal M}^2\rangle\!=\! \langle
  P_\text{spin}{\cal M}^2\rangle\!=\!\langle {\cal C}\rangle$. A telegraphic proof of $C_{{\cal C}{\cal
      B}_\text{SW}}(0)\!=\!0$ is also obtained by combining 
$\langle {\cal C}{\cal M}^2\rangle\!=\! \langle
  P_\text{spin} {\cal C}{\cal M}^2\rangle$  with
$P_\text{spin} {\cal
    C}{\cal M}^2\!=\!{\cal C} P_\text{spin} {\cal M}^2\!=\!{\cal
    C}^2$ and with Eq.~\eqref{C_CBSW}.}.

Combining $P^\text{SW}=P_\text{bond}P_\text{spin}$ with
$[P_\text{spin}]^2= P_\text{spin}$ (hence
$[P_\text{bond}P_\text{spin}]^{t>0}= [P^\text{SW}]^tP_\text{spin}$)
and with the self-adjointedness of $P_\text{spin}$ and
$P_\text{bond}$, we get for $t>0$
\begin{eqnarray}
({\cal M}^2, [P^\text{SW}]^t{\cal M}^2)&=&
({\cal M}^2, [P^\text{SW}]^t P_\text{spin}{\cal M}^2)\,,\nonumber\\
&=& ({\cal M}^2, [P^\text{SW}]^t {\cal C})\,,\nonumber\\
&=& (P_\text{spin}P_\text{bond}{\cal M}^2, [P^\text{SW}]^{t-1} {\cal C})\,,\nonumber\\
&=& ({\cal C},[P^\text{SW}]^{t-1} {\cal C})\,,\label{M2M2-LS}
\end{eqnarray}
\begin{eqnarray}
({\cal M}^2, [P^\text{SW}]^t{\cal C})&=& ({\cal C},[P^\text{SW}]^{t-1} {\cal C})\,,\label{M2C-LS}\\
({\cal C},[P^\text{SW}]^t{\cal M}^2)&=& ({\cal C},[P^\text{SW}]^t {\cal C})\,.\label{CM2-LS}
\end{eqnarray}

Now, Eqs.~(\ref{M2M2-LS},\ref{M2C-LS},\ref{CM2-LS}) tell us that
($\delta_{t,0}$ stands for Kronecker's delta, and we assume $t>0$)
\begin{eqnarray}
C_{{\cal M}^2{\cal M}^2}(t)&=&\delta_{t,0} C_{{\cal M}^2{\cal M}^2}(0)\\
&+&(1-\delta_{t,0})\, C_{{\cal C}{\cal C}}(t-1)\,,\nonumber\\
C_{{\cal C}{\cal M}^2}(t)&=&\delta_{t,0} C_{{\cal C}{\cal C}}(0)\\
&+&(1-\delta_{t,0})\,\frac{C_{{\cal C}{\cal C}}(t-1)+C_{{\cal C}{\cal C}}(t)}{2}\,.\nonumber
\end{eqnarray}
Deriving at this point Eqs.~(\ref{TILDE-R2-FINAL},\ref{rCBSW-FINAL})
is straightforward. 

We note, finally, that
\begin{equation}
({\cal C}, [P^\text{SW}]^t{\cal C})=({\cal C}, [P_\text{spin}P_\text{bond}P_\text{spin}]^t{\cal C})\,,
\end{equation}
which implies that $C_{{\cal C}{\cal C}}(t)>0$, and hence $\tau_{\text{int},{\cal C}}\geq 1/2$.

\end{document}